\documentclass[aps,prl,reprint,twocolumn,superscriptaddress,amsmath,amssymb,citeautoscript]{revtex4-1}

\usepackage{graphicx}
\usepackage{bm,amsmath,amssymb}
\usepackage{mathrsfs}
\usepackage{cases}
\usepackage{color}
 \usepackage{indentfirst}
\usepackage{epstopdf}
\usepackage{grffile}

\usepackage{adjustbox}
\usepackage{subfigure}

\DeclareGraphicsExtensions{.eps}

\newcommand{\be}{\begin{equation}}
\newcommand{\ee}{\end{equation}}
\newcommand{\bea}{\begin{eqnarray}}
\newcommand{\eea}{\end{eqnarray}}
\newcommand{\la}{\langle}
\newcommand{\ra}{\rangle}
\newcommand{\ii}{\rm i}
\newcommand{\e}{\rm e}

\newcommand{\aU}{{\uparrow}}
\newcommand{\aL}{{\leftarrow}}
\newcommand{\aR}{{\rightarrow}}
\newcommand{\aD}{{\downarrow}}
\newcommand{\aH}{{\circ}}
\newcommand{\aP}{{\bullet}}

\newcommand{\vR}{\vec{r}}
\newcommand{\vRp}{\vec{r'}}
\newcommand{\vX}{\vec{x}}
\newcommand{\vY}{\vec{y}}

\newcommand{\tr}{\rm Tr}

\renewcommand\vec{\mathbf}

\newcommand{\CState}[4]{\arraycolsep=0.5pt\def\arraystretch{0.6}\scriptsize
\begin{array}{ccc}
{#1} & {#2} & {#4} \\
 & {#3} &
\end{array}}

\newcommand{\hannover}{Institut f\"ur Theoretische Physik, Leibniz Universit\"at Hannover, Appelstr. 2, DE-30167 Hannover, Germany}

\begin{document}
\title{Deconfining disordered phase in two-dimensional quantum link models}
\author{Lorenzo Cardarelli} 
\affiliation{\hannover}
\author{Sebastian Greschner} 
\affiliation{Department of Quantum Matter Physics, University of Geneva, CH-1211 Geneva, Switzerland} 
\author{Luis Santos}
\affiliation{\hannover} 

\date{\today}
\begin{abstract}
We explore the ground-state physics of two-dimensional spin-$1/2$ $U(1)$ quantum link models, one of the simplest non-trivial lattice gauge theories 
with fermionic matter within experimental reach for quantum simulations. 
Whereas in the large mass limit we observe Ne\'el-like vortex-antivortex  and striped crystalline phases, 
for small masses there is a transition from the striped phases into a disordered phase whose properties resemble those at the Rokhsar-Kivelson point of the quantum dimer model. 
This phase is characterized on ladders by boundary Haldane-like properties, such as vanishing parity and finite string ordering.
Moreover, from studies of the string tension between gauge charges, we find that whereas the stripe phases are confined, the novel disordered phase present clear indications of being deconfined. 
Our results open exciting perspectives of studying highly non-trivial physics in quantum simulators, such as spin-liquid behavior and confinement-deconfinement transitions, 
without the need of explicitly engineering plaquette terms.
\end{abstract}
\maketitle

%%%%%%%%%%

% INTRODUCTION

Driven by tremendous progresses in the manipulation and control of ultracold quantum gases, this field is entering the era of the quantum simulation of lattice gauge theories~(LGTs)~\cite{Wilson1974}, 
with the long term goal of studying open problems of the early universe, dense neutron stars, nuclear physics or condensed-matter physics~\cite{Kogut1983, Levin2005, Wiese2013}.
Many theoretical proposals~\cite{Cirac2010,Zohar2011,Kapit2011,Zohar2012,Banerjee2012,Zohar2013,Banerjee2013,Zohar2013b,Tagliacozzo2013,Stannigel2014,Kasper2016,Cardarelli2017, Zache2018, barbiero2018} and recent seminal experiments with trapped ions~\cite{Martinez2016}, quantum dimer models in Rydberg-atoms-arrays~\cite{Bernien2017}, lattice modulation techniques~\cite{Clark2018, Goerg2019, schweizer2019}, or atomic mixtures~\cite{Mil2019} have shown first building-blocks of dynamical gauge fields and quantum link models~(QLMs), a generalization of LGT to spin-like link-variables~\cite{Chandrasekharan1997}.
However, the implementation of some building blocks of LGT, such as the ring-exchange corresponding to magnetic field dynamics in analogue implementations of quantum electrodynamics, 
require further theoretical and experimental breakthroughs, although there has been progress on isolated plaquettes~\cite{dai2017four} and recent promising proposals~\cite{Celi2019, Bohrdt2019}.
% Here there is also a paper of Marcello (Surace et al) on the re-interpretation of Rydberg experiments. We should cite that.

In this paper we show how already the simplest mid-term experimental realizations, without plaquette terms, may be able to explore a wide area of non-trivial phenomena of LGTs. In particular, we report in this Letter that 
the two-dimensional~(2D) QLM is characterized by the emergence of a quantum phase transition between confined crystalline phases and an exotic deconfined disordered phase with certain resemblance to Rokshar-Kivelson~(RK) states~\cite{rokhsar1988} or resonating valence bond liquids~\cite{Shastry1981,Anderson1987, Moessner2002, Ralko2005}. Hence, these relatively simple systems provide a pristine test-bed for 
the study of highly non-trivial physics, such as spin liquids, confinement-deconfinement transitions, and exotic dynamical or thermalization properties~\cite{Bernien2017, Brenes2018, feldmeier2019emergent}, such as the formation of quantum many-body scars in constrained systems~\cite{turner2018scars} and their fundamental link to confinement. Interestingly, QLMs may be experimentally realized in quantum gases within the next years. 
Whereas several proposals using Fermi-Bose mixtures have been reported~\cite{Banerjee2012,Zohar2012,Banerjee2013,Stannigel2014,Kasper2016,Zache2018,Mil2019}, we recently discussed~\cite{Cardarelli2017} a minimalistic realization of QLMs with a single fermionic species 
that simulates the spin-$1/2$ links using multi-orbital physics in optical superlattices~\cite{Wirth2011}. We note, however, that the latter may be analogously replaced by hyper-fine or spatial degrees of freedom allowing for a large flexibility of the proposal.

%%%%%%%%%%%%

% FIGURE 1

\begin{figure}[t]
\centering
\includegraphics[width=1.\linewidth]{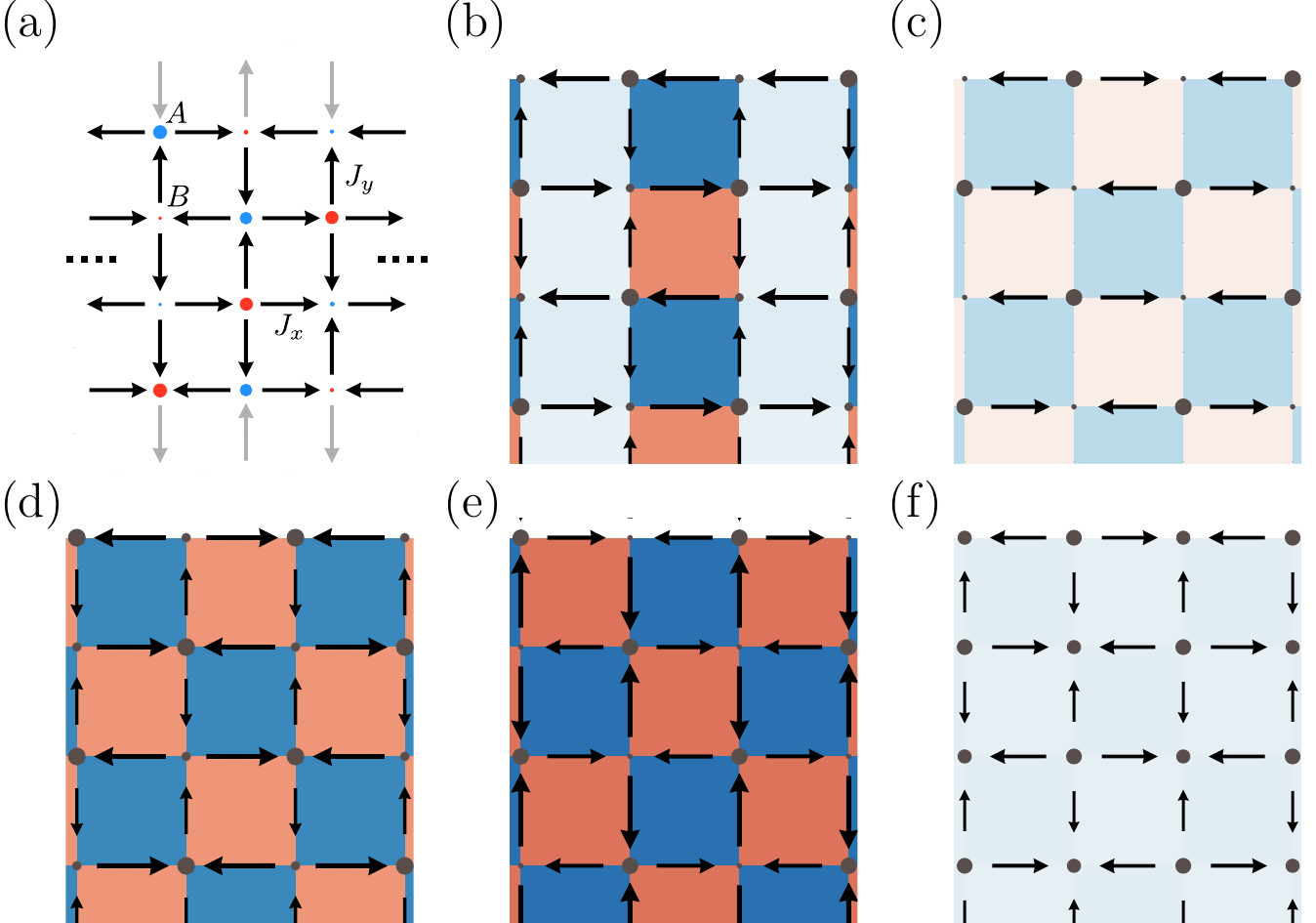}
\caption{(a) Sketch of the QLM model on a square ladder~(gray spins depict staggered boundary conditions) or on a torus (associating opposite gray spins at the two edges). 
(b-f) Sketches of the phases discussed in the text: for $\mu=0.8$ striped phases (b) $J_y=2.4 J_x$~(Sy), (c) $J_y=0.2 J_x$~(Sx); 
for $\mu=-0.8$ vortex-antivortex phases (d) $J_y=2.4 J_x$~(VA)  (e) $J_y=0.2 J_x$~(VA$'$), and the (f) disordered/deconfined D phase for $\mu=0$ and $J_x=J_y$. The size of the bullets depicts 
$\la n_{\vR}\ra$, the arrow size $\la S_{\vR,\vRp}^z\ra$, and the plaquette colors the vorticity $\la Q_{\vR}\ra$. The arrow sizes of (f) has been scaled up by a factor of $2$ for clarity.}
\label{fig:sketch}
\end{figure}

%%%%%%%%%%%%

%%%%%%%%%%%%

% FIGURE 2
% We should change the order of the figures a <-> b, and c <-> d

\begin{figure*}[t]
\centering
\includegraphics[width=0.99\linewidth]{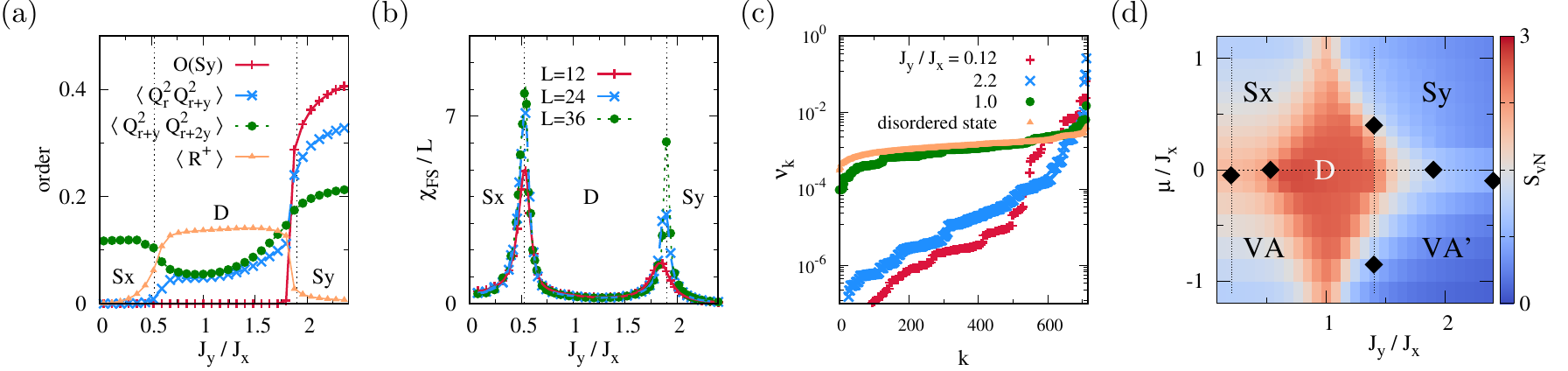}
\caption{Four-leg cylinder at $\mu=0$: 
(a) Nearest and next-nearest neighbor flippability correlations, staggered flippability $O(Sy) = \sum_\vR (-)^x \la Q_{\vR}^2 \ra$, as well as the expectation value of the ring-exchange $\la R^+_{\vR} \ra$.
(b) Fidelity susceptibility with $L=12$, $24$ and $36$ rungs.
(c) Local Hilbert space distribution $\nu_k$ of the central rung~(see text).
(d) Sketched phase diagram of the QLM. Color codes depict the von-Neumann bipartite entanglement entropy $S_{vN}$ of the central rung. 
Points depicts the estimated phase transition points, by extrapolating the peak positions of the fidelity susceptibility evaluated along the cuts indicated by the dotted lines (DMRG data).}
\label{fig:4t_cuts_mu0}
\end{figure*}

%%%%%%%%%%%%

%%%%%%%%%%%%%%%%%%%%%

% 2D QLM

\paragraph{2D QLM.--} We consider in the following a QLM on a square lattice described by the Hamiltonian
\begin{align}
H =  - \sum_{\vR \vRp} J_{\vR\vRp} \left( \psi_\vR^\dagger S^+_{\vR\vRp} \psi_{\vRp} +  {\rm h.c.} \right)
+ \sum_\vR \mu_\vR n_\vR, 
\label{eq:QLM}
\end{align}
% mu_r=mu for A and -mu for B. For mu>>0 piling on B, this means that B goes with -mu
where $\psi_\vR^\dagger$ is the fermionic operator at site $\vR$, and $n_\vR = \psi_\vR^\dagger \psi_\vR$. 
In our case the gauge field is given by spin-$1/2$ operators $S^+_{\vR\vRp}$ placed at the link between sites $\vR$ and $\vRp$. The 
amplitudes $J_{\vR,\vR\pm\vX}=J_x$ and $J_{\vR,\vR\pm\vY}=J_y$ characterize, respectively, the hops along the $x$ and $y$ directions~(Fig.~\ref{fig:sketch}~(a)). 
We enforce at each site the Gauss law $[H, G_\vR]=0$ with $G_\vR = \varepsilon_\vR - n_\vR + \sum_{\vec{k} \in \vX,\vY} ( S^z_{\vR,\vR+\vec{k}} - S^z_{\vR,\vR-\vec{k}} )$,
with $\varepsilon_{\vR\in A}=1$ and $\varepsilon_{\vR\in B}=0$. 
The staggered potential, $\mu_{\vR\in A}=\mu$ and $\mu_{\vR\in B}=-\mu$ can be interpreted as the mass of particles on B sites and anti-particles on A sites. 

We focus on the ground-states of the QLM at half fermion filling and gauge vacuum~($G_\vR=0$) on square ladders and cylinders. We study up to $L_x=100$ rungs and $L_y=4$ legs 
by means of density matrix renormalization group~(DMRG) techniques~\cite{Schollwock2011} adapted to the local gauge symmetry~\cite{Dalmonte2016cont, Tschirsich2019, Chepiga2019dmrg, DMRGnote}.
We introduce at this point the ring-exchange operators $R^+_\vR= S_{\vR,\vR+\vX}^+ S_{\vR+\vX,\vR+\vX+\vY}^- S_{\vR+\vX+\vY,\vR+\vY}^- S_{\vR+\vY,\vR}^+$ and $R_\vR^-=(R_\vR^+)^\dagger$.
These operators characterize plaquette states: $R_\vR^-$~($R_\vR^+$) flips a vortex~(antivortex) into an antivortex~(vortex), being zero otherwise, 
and $Q_{\vR}=(R_\vR^+R_\vR^- - R_\vR^-R_\vR^+)=1$~(-1) for vortex~(antivortex) and $Q_{\vR}=0$ otherwise~\cite{Arrows}.
%, and  $R_\vR^2 = Q_\vR^2$ is $1$~($0$) for flippable~(unflippable) plaquettes. 
%Note, however, that in contrast to QED realizations or to the quantum dimer model~(QDM), they do not appear explicitly in the Hamiltonian~\eqref{eq:QLM}.

%%%%%%%%%%%%%%%%%%%%%%%%%

% LARGE MASS

{\em Large mass limit.--} First insights are obtained from the limit $|\mu| \gg J_{x,y}$, which, in contrast to the two-leg QLM~\cite{supmat}~\nocite{aklt, Pollmann2010, Pollmann2012, Pollmann2012A}, is different 
for $\mu>0$ and $\mu<0$. For $\mu>0$, particles 
are pinned in B sites~(Figs.~\ref{fig:sketch}~(b) and (c)). 
Local states are characterized by the expectation value of two spin-$1$ operators, $S^z_{\vec{k}}(\vR) = S^z(\vR-\vec{k}, \vR) + S^z(v,\vR+\vec{k})$, with $\vec{k}=\vX,\vY$. 
For $J_x<J_y$, second-order terms select a ground-state manifold of two states with $S^z_y(\vR)=0$. Fourth-order ring-exchange $\propto J_x^2 J_y^2 / |\mu|^3$~\cite{Cardarelli2017} 
favors a configuration of columns of flippable vortex-antivortex and non-flippable plaquettes~(Fig.~\ref{fig:sketch}~(b)). We denote this striped phase Sy. 
In a 2D model, a corresponding striped phase Sx of alternating flippable and non-flippable rows of plaquettes is expected for $J_y<J_x$. 
However, on the finite-size cylinders we study,
%the unbroken translational symmetry
the translational symmetry along the \textit{y} direction
results in blurred spin averages~(Fig.~\ref{fig:sketch}~(c)). 
Correlations of the flippability operators reveal the Sx character~(Fig.~\ref{fig:4t_cuts_mu0}~(a)):  whereas $\la Q_{\vR}^2 Q_{\vR+\vY}^2 \ra$ vanishes, $\la Q_{\vR}^2 Q_{\vR+2\vY}^2 \ra$ remains finite.
Staggered boundary spins stabilize the Sx ordering in ladders~\cite{supmat}.

For large negative mass, $-\mu \gg J_{x,y}$, particles are pinned to the A-sites, reducing the local Hilbert space to a six-dimensional manifold of spin configurations satisfying Gauss' law.
Second-order processes favor states with $\la S^z_{\mathbf{x}}(\vR) \ra=\la S^z_{\mathbf{y}}(\vR) \ra=0$, leading to a checkerboard ground-state pattern of vortex-antivortex~(VA) plaquettes~(Figs.~\ref{fig:sketch}~(d) and (e)). For $J_x<J_y$ we dub this phase VA, and VA$'$ for $J_y<J_x$. These two phases are uniquely defined and do not exhibit any spontaneously broken translational symmetry like in a Ne\'el-like phase.

%%%%%%%%%%%%%%%%%%%

{\em Emerging disordered phase.--}  At low $\mu$ particle fluctuations become important, leading to a particularly intriguing physics.
For $\mu\sim 0$ we observe three distinct phases as a function of $J_y/J_x$, as can be seen in Fig.~\ref{fig:4t_cuts_mu0}~(b) for the four-leg cylinder 
by the distinct diverging peaks in the fidelity susceptibility $\chi_{FS} = \lim_{J_y-J_y' \to 0} \frac{-2 \ln |\langle \Psi_0(J_y) |\Psi_0(J_y') \rangle| }{(J_y-J_y')^2}$,
where $|\Psi_0\rangle$ is the ground-state wave-function. We observe a similar behavior for three- and four-leg ladders~\cite{supmat}. 
Whereas for $\mu=0$ for $J_y \ll J_x$~($J_y \gg J_x$) the system is in the Sx~(Sy) phase, for $J_x\sim J_y$ an intermediate gapped phase occurs characterized by vanishing 
$\la Q_{\vR} \ra$ and $\la Q_{\vR}^2 \ra$, but a large expectation value of the ring-exchange $\la R^+_{\vR} \ra$.

A crucial insight on the physics of the intermediate phase is provided by the analysis of the reduced density matrix $\rho_c = \tr |\Psi_0\ra\la \Psi_0|$ for the central rung~(where the trace runs over all other rungs) 
in the~(Fock-like) eigenbasis $\phi_k$ of $S^z_{\vR\vRp}$ and $n_{\vR}$.
In Fig.~\ref{fig:4t_cuts_mu0}~(c) we show its diagonal elements $\nu_k = \la \phi_k | \rho_c | \phi_k \ra$, an effective local Hilbert space distribution, sorted by amplitude, for the case of a four-leg cylinder. The Sx and Sy phases are strongly localized in Fock space, i.e. $\nu_k$ has most weight for few basis states. The intermediate phase, however, exhibits a drastically different, much flatter distribution, where many local Fock states contribute with similar weight. The disordered character of the intermediate phase is also witnessed by the entanglement entropy $S_{vN}=-\tr(\rho_c \ln \rho_c)$, which we depict in Fig.~\ref{fig:4t_cuts_mu0}~(d).

The intermediate phase thus closely resembles the Rokhsar-Kivelson~(RK) point, which contains an equal superposition of all dynamically connected states. We also show in Fig.~\ref{fig:4t_cuts_mu0}~(c) the corresponding distribution of  $\nu_k$ for a classical RK state, which compares well to the ground state obtained by the DMRG simulation. We estimate the overlap between the two states to 
be $0.97$~(see~\cite{supmat} for a more detailed comparison between the DMRG simulation of the intermediate phase and the classical RK state, which also reproduces the spin and density configuration pattern of Fig.~\ref{fig:sketch}~(f)). We hence characterize the intermediate gapped phase as a disordered~(D) phase. Note, that due to the different Gauss' law on A and B sites, this phase still exhibits a slight particle imbalance between A and B sites, as well as finite link-variable expectation values, as shown in Fig.~\ref{fig:sketch}~(f).

%%%%%%%%%

% FIGURE 3

\begin{figure}[tb]
\begin{minipage}[t]{.03\linewidth}
\raisebox{2.8cm}[0cm][0cm]{(a)}
\end{minipage}
\begin{minipage}[t]{.45\linewidth}
	\includegraphics[height=2.7cm]{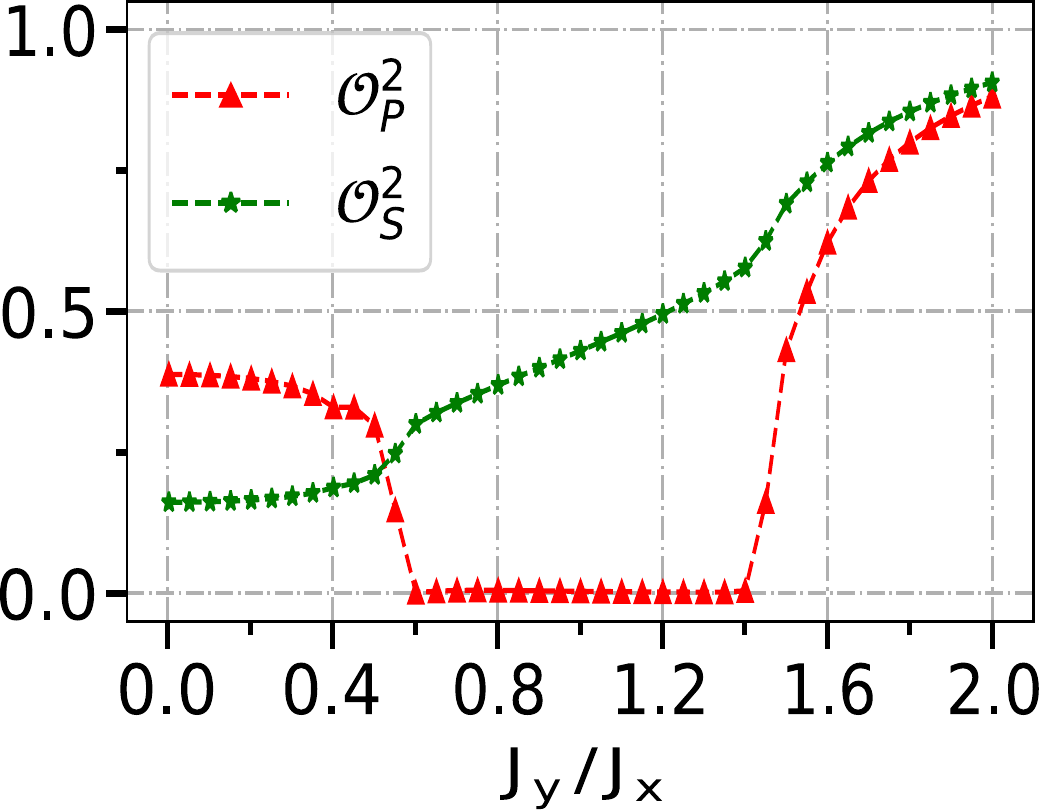}
\end{minipage}
\begin{minipage}[t]{.03\linewidth}
\raisebox{2.8cm}[0cm][0cm]{(b)}
\end{minipage}
\begin{minipage}[t]{.45\linewidth}
	\includegraphics[height=2.7cm]{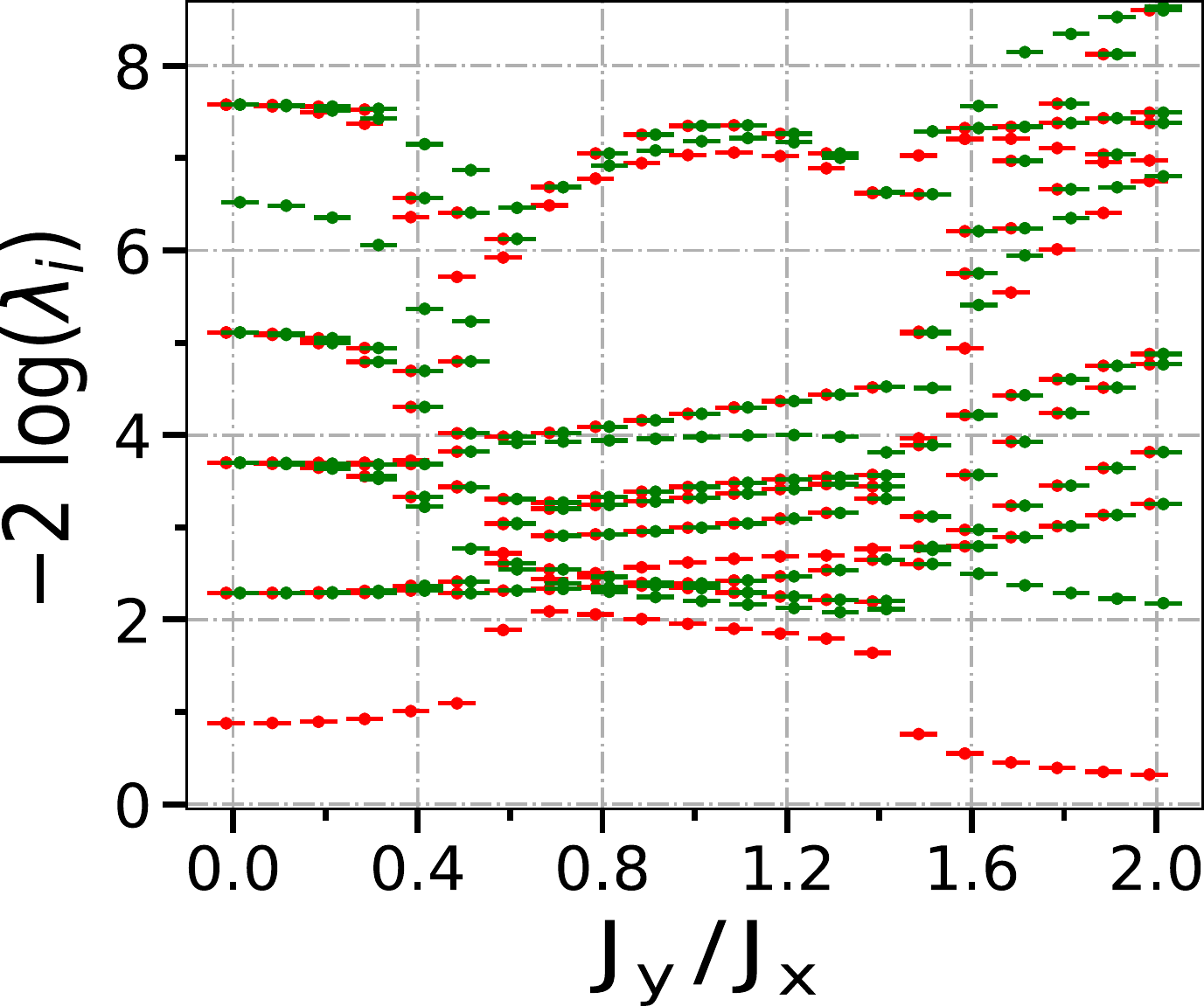}
\end{minipage}
\caption{(a) Parity and string order along the boundary legs of a four-leg ladder, obtained using DMRG for $L_x=100$ rungs and $\mu=0$; (b) largest eigenvalues of the entanglement spectrum, obtained 
after dividing the system into two parts along the central rung.}
\label{fig:entanglement_SO}
\end{figure}

%%%%%%%%%

%%%%%%%%%%%%%%%%

% EDGE HALDANE ORDER

\paragraph{Edge Haldane order.--} We focus at this point on the edges of a QLM ladder, where the physics can be well understood from 
a mean-field-like strongly simplified 1D model in which we fix the upper boundary links for each site in a staggered configuration, and 
allow the lower spins to fluctuate with an amplitude $J_y$. Six local states are possible:
$\left|0_\pm \right\rangle = ( \left| \CState{\aL}{\aH}{\aU}{\aL} \right\rangle \pm \left| \CState{\aL}{\aP}{\aD}{\aL} \right\rangle ) /\sqrt2$,
$\left|\tilde{0}_\pm\right\rangle = ( \left| \CState{\aR}{\aH}{\aU}{\aR} \right\rangle \pm \left| \CState{\aR}{\aP}{\aD}{\aR} \right\rangle) /\sqrt2$,
$\left|\alpha \right\rangle = \left| \CState{\aR}{\aH}{\aD}{\aL} \right\rangle$, and 
$\left|\beta \right\rangle = \left| \CState{\aL}{\aP}{\aU}{\aR} \right\rangle$.
Gauss' law imposes further restrictions on the allowed sequence of these local states: $0$ may be followed at its right by $0$ or $\beta$~($0\to 0,\beta$); $\tilde{0} \to \tilde{0}, \alpha$; $\alpha \to \beta, 0$; and 
$\beta \to \alpha, \tilde{0}$~(we remove the $\pm$ index). By construction, Gauss' law enforces a Ne\'el-like order of $\alpha$ and $\beta$ states diluted by 
an arbitrary number of intermediate $0$ or $\tilde{0}$ states.
The model Hamiltonian, given by
\begin{align}
H_{\rm 1D} =  - J_x \sum_{x} \psi_{x}^\dagger S_{x,x+1}^+ \psi_{x} - J_y \sum_{x} \psi_{x}^\dagger S_{x}^+ + {\rm H.c.}
\label{eq:QLM1D}
\end{align}
exhibits three ground-state phases (here we neglect a staggered potential term). For $J_y\ll J_x$ the ground state is $\cdots |\alpha\ra |\beta\ra |\alpha\ra |\beta\ra \cdots$, whereas for $J_y\gg J_x$ 
the states $\cdots |0_-\ra |0_-\ra \cdots$ and $\cdots |\tilde{0}_-\ra |\tilde{0}_-\ra \cdots$ have the lowest energy. Interestingly, for $J_x\sim J_y$ an intermediate phase with Haldane-like diluted Ne\'el order emerges, 
that resembles the SPT phase of Ref.~\cite{Cardarelli2017}.
We may describe this intermediate phase by a minimal AKLT-like~\cite{aklt} state with a two-fold degenerate entanglement spectrum and a non-vanishing string order ${\cal O}_{S}^2 = \lim_{|x-x'|\to\infty} \la S^z_x \e^{\ii \pi \sum_{x<k<{x'}} S^z_k} S^z_{x'} \ra$, while parity order ${\cal O}_{P}^2 = \lim_{|x-x'|\to\infty} \la \e^{\ii \pi \sum_{x<k<{x'}} S^z_k} \ra$ is exponentially suppressed~\cite{supmat}.

While being a drastically simplified description, it captures essential ingredients of ladder QLMs. In particular, fixing in a ladder the boundary spins to a staggered configuration enforces the dilute Ne\'el order on the boundary leg. We, hence, plot in Fig.~\ref{fig:entanglement_SO}~(a) the string- and parity order measured along the boundary leg of a 4-leg ladder. Indeed the D phase is characterized by a finite string order and a vanishing parity order, resembling closely the SPT phase discussed for the above mentioned 1D model or the two-leg QLM of Ref.~\cite{Cardarelli2017}. However, for $L_y>2$ the parity order remains finite in the D phase if measured on the inner legs, and
the phase is not topological. The entanglement spectrum is no longer strictly two-fold degenerate. Interestingly, however, we observe a robust gap in the entanglement spectrum of the D phase between a low-lying manifold and the rest.

%%%%%%%%%

% FIGURE 4
\begin{figure}[tb]
\centering
\includegraphics[width=1.\linewidth]{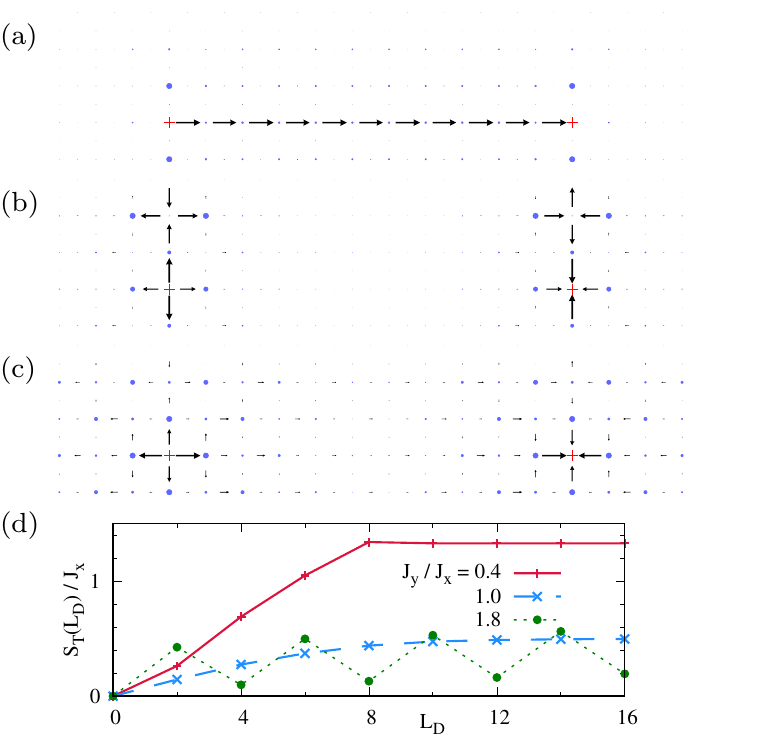}
\caption{(a) Average fermionic density and bond configuration of two charges at a distance of $L_D=12$ sites, after substracting the charge-free configuration. We employ DMRG for a cylinder with $L_y=4$ legs and $L=36$ rungs and $\mu=0.4 J_x$. (a) Sy phase~($J_y=1.8 J_x$), (b) Sx phase~($J_y=0.4 J_x$), (c) D phase~($J_y=J_x$). (d) String tension $S_T$ as a function of the distance between the defects $L_D$.}
\label{fig:config_4T_ST}
\end{figure}

%%%%%%%%%

%%%%%%%%%%%%%%%%%%

% STRING TENSION

\paragraph{String tension.--} Finally, we discuss the properties of gauge charges on top of the vacuum state.
We insert two charges by locally adjusting Gauss' law to $G_\vR = \pm 1$ on two sites separated by a distance $L_D$ in $x$-direction, 
and study the string formation for the case of a four-leg cylinder.
Example configurations are shown in Figs.~\ref{fig:config_4T_ST}~(a)-(c) for Sy, Sx and D phases after subtracting the spin and fermion configuration of the charge-free system.
Comparing the energy $E(L_D)$ with the energy $E_0$ of the charge-free state, we obtain the string tension, $S_T(L_D) = E(L_D)-E_0$~(Fig.~\ref{fig:config_4T_ST}~(d)), 
which characterizes the confining properties~\cite{Tschirsich2019}.

Only Sy shows a clear string formation~(Fig.~\ref{fig:config_4T_ST}~(a)). The tension increases linearly 
in a staggered way due to the broken symmetry, as depicted in Fig.~\ref{fig:config_4T_ST}(d). This is a clear signature of the confinement of excitations. 
For the Sx phase the increase of potential energy is also linear and very large compared to the other phases. Here, however, after some distance the string breaks and is wrapped around the cylinder in $y$-direction (see Fig.~\ref{fig:config_4T_ST}~(b)). Also the string tension flattens after this point as shown in Fig.~\ref{fig:config_4T_ST}~(d) for $J_y/J_x=0.4$. Interestingly, the two charges become, hence, effectively deconfined due to the finite size of the system in y-direction.

In the D phase the tension grows slowly with $L_D$ and potentially finally saturates, indicating charge deconfinement.  Contrary to the Sx phase we observe in Fig.~\ref{fig:config_4T_ST}~(c) the formation of a symmetric broad but localized perturbation of the spin and charge background around the defects. 
Even though due to the limited system size we cannot distinguish the saturation of the string tension from a further slow~(e.g. logarithmic) growth, these results show that Sx, Sy and D phases exhibit drastically different confinement and deconfinement properties.

%%%%%%%%%%%%%%

% CONCLUSIONS

\paragraph{Conclusions.--} We studied the ground state of a 2D spin-$1/2$ QLM, which may be realizable in quantum gas lattice gauge simulators in the foreseeable future. 
Despite the absence of plaquette terms, 2D QLMs are characterized by a highly nontrivial physics.
%Even though key elements of true LGT, may be missing, a rich physics can be observed in the vacuum state by controlling the ratio of the hopping amplitudes in the $x$- and $y$-direction, and the staggered potential fermion mass-like term. 
As a main result, we have found an emergent deconfined disordered phase for $\mu\sim 0$ and $J_x\sim J_y$, which closely resembles an RK phase. On finite ladder systems with staggered boundary spins this phase exhibits Haldane-like ordering at the edge legs. While being limited to small transversal lengths $L_y\leq 4$, the observed features qualitatively remain robust over two-, three- and four-leg ladders and four-leg cylinders, strongly hinting that the intermediate disordered phase may survive in more general 2D lattices, which might inspire further numerical efforts in this direction.

Our results open the interesting possibility to study a wealth of phenomena such as deconfinement-confinement transition and RVB-like physics in quantum gas lattice gauge simulators, without 
the need of explicitly realizing ring-exchange and RK terms. The dynamics of these systems may be particularly interesting. Further experimental and theoretical studies should reveal the potentially unconventional thermalization properties~\cite{turner2018scars, feldmeier2019emergent} of constrained systems with fermionic matter.

%%%%%%%%%%%%%%%%%%%%%%%%%%%%%%%%%%%%%%%%%%%%%%%%%%%%%%%%%%%%%%%%%%%%%%%%%%%%%%%%%%%%%%%%%%
\begin{acknowledgments}
We thank Alessio Celi and Marcello Dalmonte for enlightening discussions. S.~G. acknowledges support by the Swiss National Science Foundation under Division II. 
L.~C. and L.~S. thank the support of the German Research
Foundation Deutsche Forschungsgemeinschaft (Project No. SA 1031/10-1 and the Excellence Cluster QuantumFrontiers).
Simulations were carried out on the cluster system at the Leibniz University of Hannover and the boabab cluster of the University of Geneva.
\end{acknowledgments}

\bibliography{references}

\end{document}

\title{Supplementary material to "Deconfining disordered phase in two-dimensional quantum link models"}

\author{Lorenzo Cardarelli} 
\affiliation{\hannover}
\author{Sebastian Greschner} 
\affiliation{Department of Quantum Matter Physics, University of Geneva, CH-1211 Geneva, Switzerland} 
\author{Luis Santos}
\affiliation{\hannover} 

\date{\today}
\begin{abstract}
In this supplementary material we give details on our numerical results for the simplified 1D model, two-, three- and four-leg ladders. We also show additional data for the classical RK-states and the calculation of the string tension.
\end{abstract}
\maketitle

%%%%%%%%%%%%%%%%%%%%%%%%%%%%%%%%%%%%%%%%%%%%%%%%%%%%%%%%%%%%%%%%%%%%%%%%%%%%%%%%%%%%%%%%%%
\section{Simplified 1D model}

In the following, we present some details on the simplified 1D model described in Eq.~\eqQLMoneD{} of the main text. As discussed in the main text, the model can be described in a basis of six local states:
\begin{align*}
\left|0_\pm \right\rangle &= ( \left| \CState{\aL}{\aH}{\aU}{\aL} \right\rangle \pm \left| \CState{\aL}{\aP}{\aD}{\aL} \right\rangle ) /\sqrt2  \,, \\
\left|\tilde{0}_\pm\right\rangle &= ( \left| \CState{\aR}{\aH}{\aU}{\aR} \right\rangle \pm \left| \CState{\aR}{\aP}{\aD}{\aR} \right\rangle) /\sqrt2  \,, \\
\left|\alpha \right\rangle &= \left| \CState{\aR}{\aH}{\aD}{\aL} \right\rangle \,, \\
\left|\beta \right\rangle &= \left| \CState{\aL}{\aP}{\aU}{\aR} \right\rangle
\end{align*}
with the restrictions on the allowed sequence of these local states by Gauss' law: $0$ may be followed at its right by $0$ or $\beta$~($0\to 0,\beta$); $\tilde{0} \to \tilde{0}, \alpha$; $\alpha \to \beta, 0$; and 
$\beta \to \alpha, \tilde{0}$~(we remove the $\pm$ index).
As mentioned in the main text the ground states can be understood from a $J_y\ll J_x$ and $J_y\gg J_x$ limit as $\cdots |\alpha\ra |\beta\ra |\alpha\ra |\beta\ra \cdots$ and $\cdots |0_-\ra |0_-\ra \cdots$ or $\cdots |\tilde{0}_-\ra |\tilde{0}_-\ra \cdots$ crystalline states respectively.

\begin{figure}[tb]
\includegraphics[width=\linewidth]{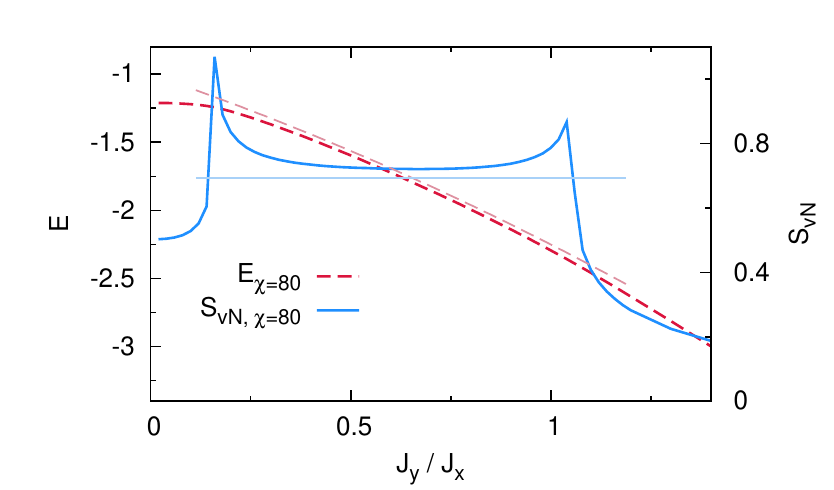}
\caption{Ground-state energy per site $E / L$ and average entanglement entropy $S_{vN}$ of the simplified 1D model~(infinite DMRG simulation, $\chi=80$). The lighter colored curves depict the $\chi=2$ ansatz described in the main text and Eq.~\eqref{eq:lgt1dMPS}.}
\label{fig:S_1d_m0}
\end{figure}

To shed light into the properties of the intermediate Haldane-like phase, we define the variational MPS ground state as $|\Psi\rangle = \sum_\sigma \Lambda \Gamma_\sigma | \sigma \rangle$. The $\Gamma_\sigma$ matrices, considered as an automorphism of $\chi\times\chi$ matrices, fulfill the following simple algebraic relations defined by Gauss' law.
A simple example is given by the following matrices, with the lowest non-trivial bond dimension $\chi=2$:
\begin{align}
\Gamma_\alpha &=  \sqrt{2} \left(
\begin{array}{cc} 
0 & 0 \\
\sin\phi & 0 
\end{array}\right) \nonumber\\
\Gamma_\beta &= \Gamma_A^T \nonumber\\
\Gamma_{\tilde{0}_\pm} &= \left(
\begin{array}{cc} 
\cos\phi & 0 \nonumber\\
0 & 0 
\end{array}\right) \\
\Gamma_{0_\pm} &= \left(
\begin{array}{cc} 
0 & 0 \\
0 & \cos\phi 
\end{array}\right)
\label{eq:lgt1dMPS}
\end{align}
and $\Lambda=\{ \frac{1}{\sqrt{2}}, \frac{1}{\sqrt{2}} \}$ which resembles an AKLT-model like state~\cite{aklt}. This ansatz has minimal energy for $\phi_{min}=\arctan\left(\frac{2J_x-J_y}{2J_x + J_y}\right)$. With this $E_{min}=-\frac{(2J_x + J_y)^2}{4J_x}$, which gives a reasonably good approximation of the actual ground-state energy in this regime obtained by a DMRG simulation, as shown in Fig.~\ref{fig:S_1d_m0}. Also the entanglement entropy of $\log(2)$ describes already well the actual values obtained by DMRG simulations with a higher bond-dimension~(see Fig.~\ref{fig:S_1d_m0}). 
The string order is given by $O_{SO} = \lim_{|x-x'|\to\infty} \la S^z_x \e^{\ii \pi \sum_{x<k<{x'}} S^z_k} S^z_{x'} \ra = (-1)^{i-j} \frac{\cos^4\phi}{4}$, while the parity order $O_{PO} = \lim_{|x-x'|\to\infty} \la \e^{\ii \pi \sum_{x<k<{x'}} S^z_k} \ra$ is exponentially suppressed.

\begin{figure*}[tb]
\begin{minipage}[t]{.03\linewidth}
\raisebox{3.8cm}[0cm][0cm]{(a)}
\end{minipage}
\begin{minipage}[t]{.26\linewidth}
	\includegraphics[height=4cm]{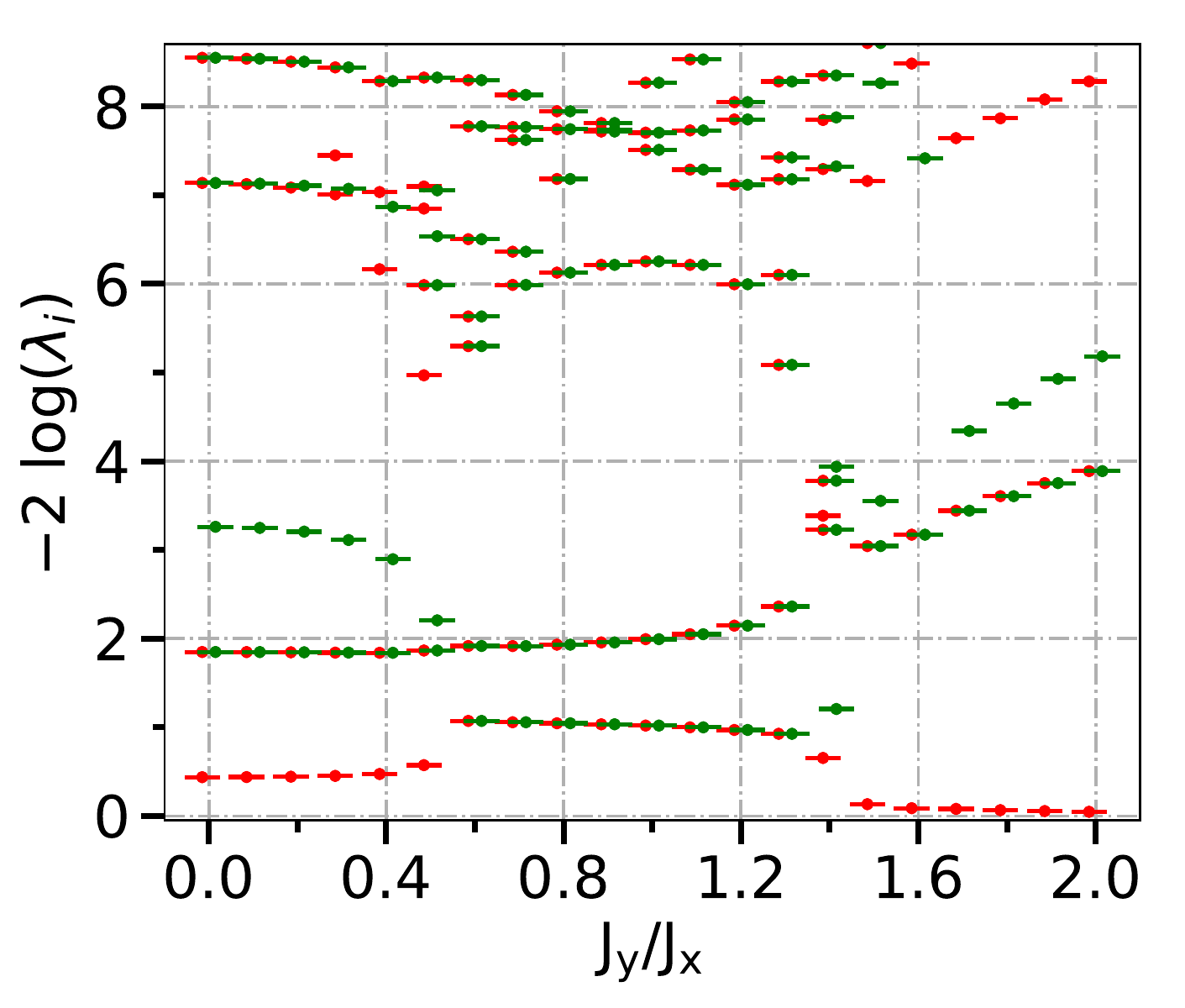}
\end{minipage}
\begin{minipage}[t]{.03\linewidth}
\raisebox{3.8cm}[0cm][0cm]{(b)}
\end{minipage}
\begin{minipage}[t]{.26\linewidth}
	\includegraphics[height=4cm]{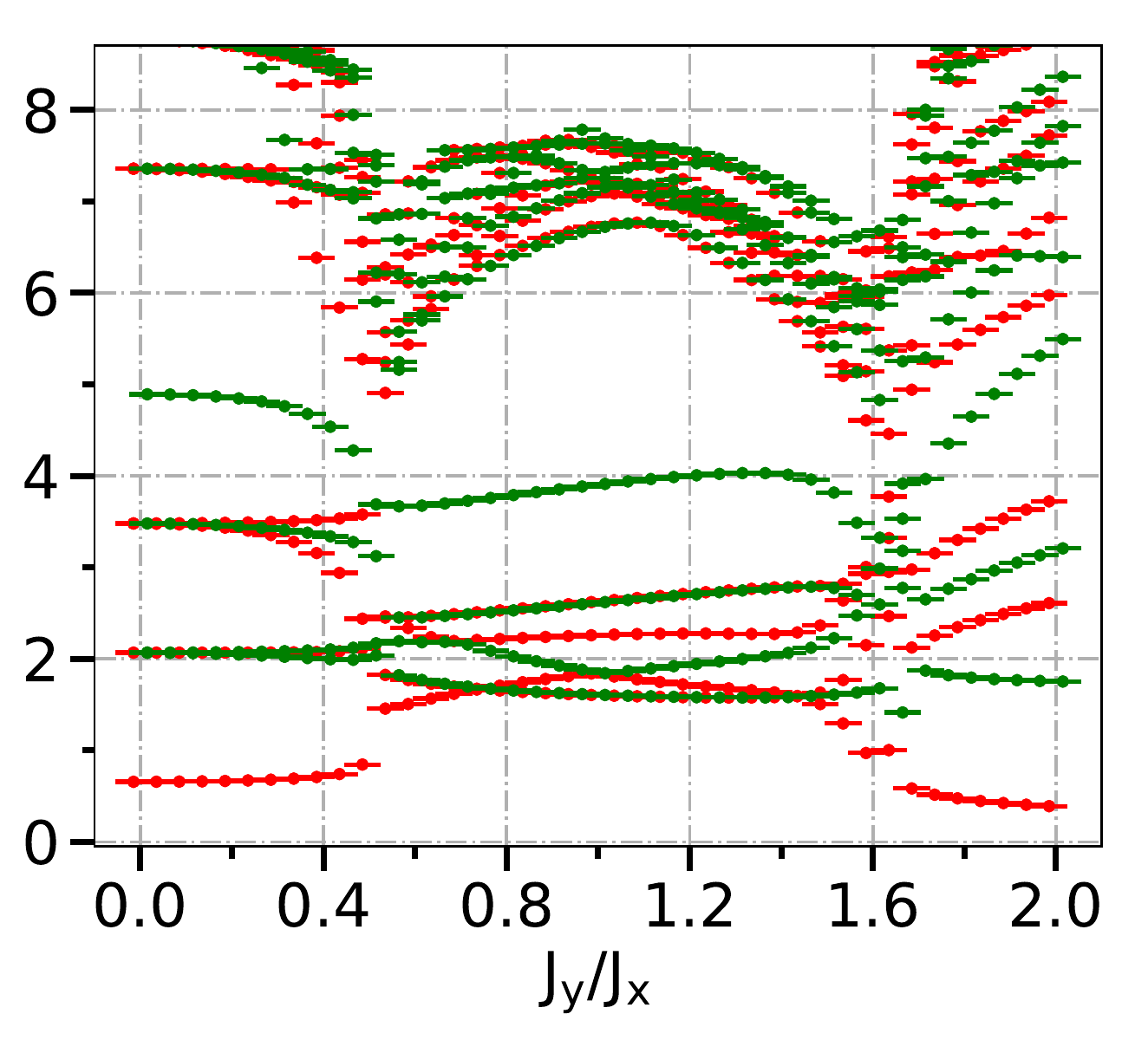}
\end{minipage}
\begin{minipage}[t]{.03\linewidth}
\raisebox{3.8cm}[0cm][0cm]{(c)}
\end{minipage}
\begin{minipage}[t]{.26\linewidth}
	\includegraphics[height=4cm]{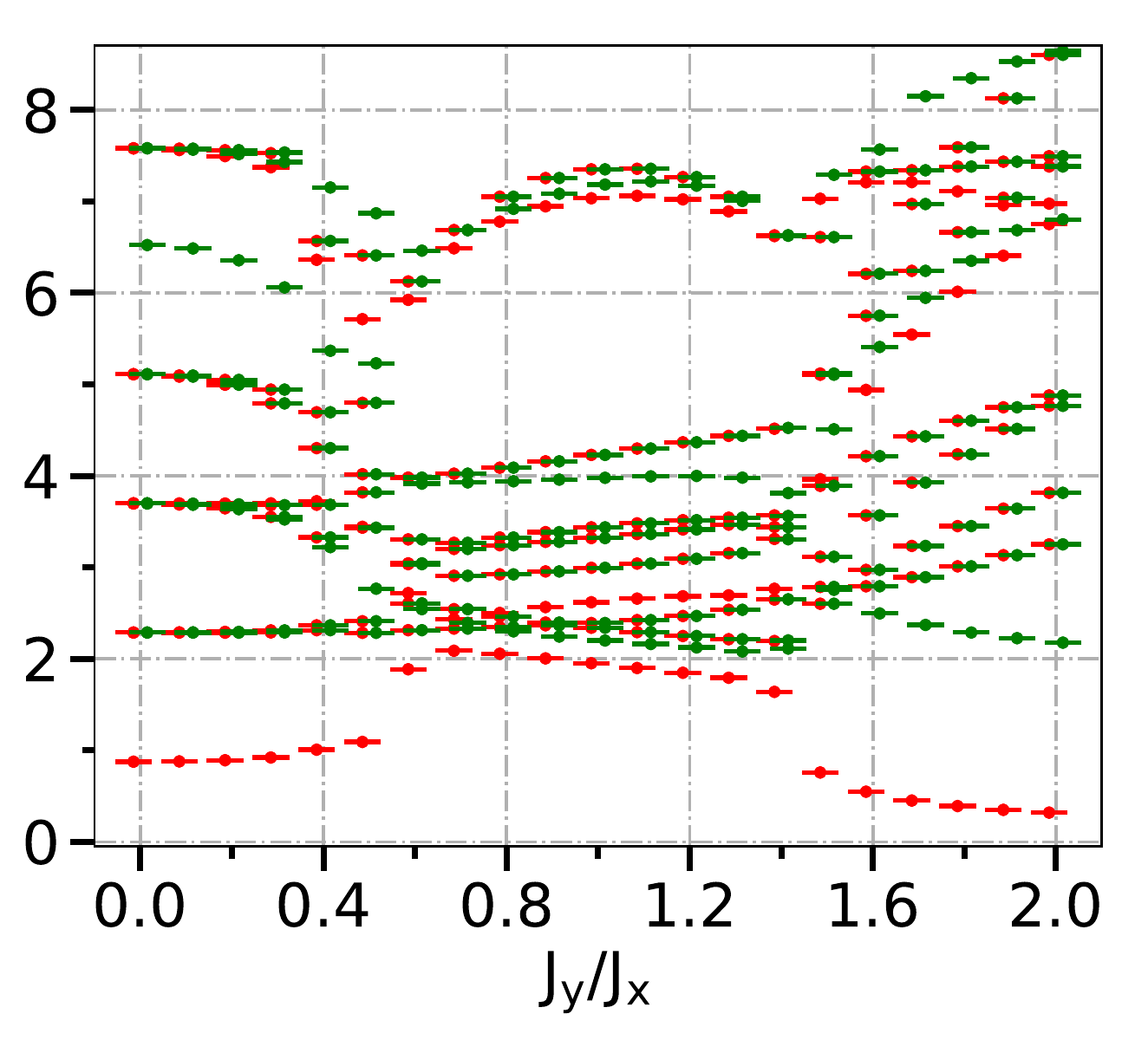}
\end{minipage}
\caption{Mid-system entanglement spectrum of the QLM Eq.~(1) of the main text at $\mu=0$ for (a) two-leg ladder, (b) three-leg ladder, (c) four-leg ladder, with $L=100$ and $\chi=100$ (DMRG data).}
\label{fig:S_entanglement}
\end{figure*}
%%

%%%%%%%%%%%%%%%%%%%%%%%%%%%%%%%%%%%%%%%%%%%%%%%%%%%%%%%%%%%%%%%%%%%%%%%%%%%%%%%%%%%%%%%%%%
\section{2-leg ladder}

Let us briefly review the results obtained for Eq.~\eqQLM{} of the main text for the two-legs ladder system~(further details of the two-leg ladder model are discussed in Ref.~\cite{Cardarelli2017}). For staggered boundary links, the model becomes symmetric in the mass term $\mu = -\mu$ (note that, as mentioned in the main text, this is not the case for systems with more than two legs). In the ground-state phase diagram three different phases emerge corresponding to the phases obtained in the simplified 1D model of the previous section.

\begin{figure}[tb]
\begin{minipage}[t]{.05\linewidth}
\raisebox{3.1cm}[0cm][0cm]{(a)}
\end{minipage}
\begin{minipage}[t]{.7\linewidth}
	\includegraphics[width=.99\linewidth]{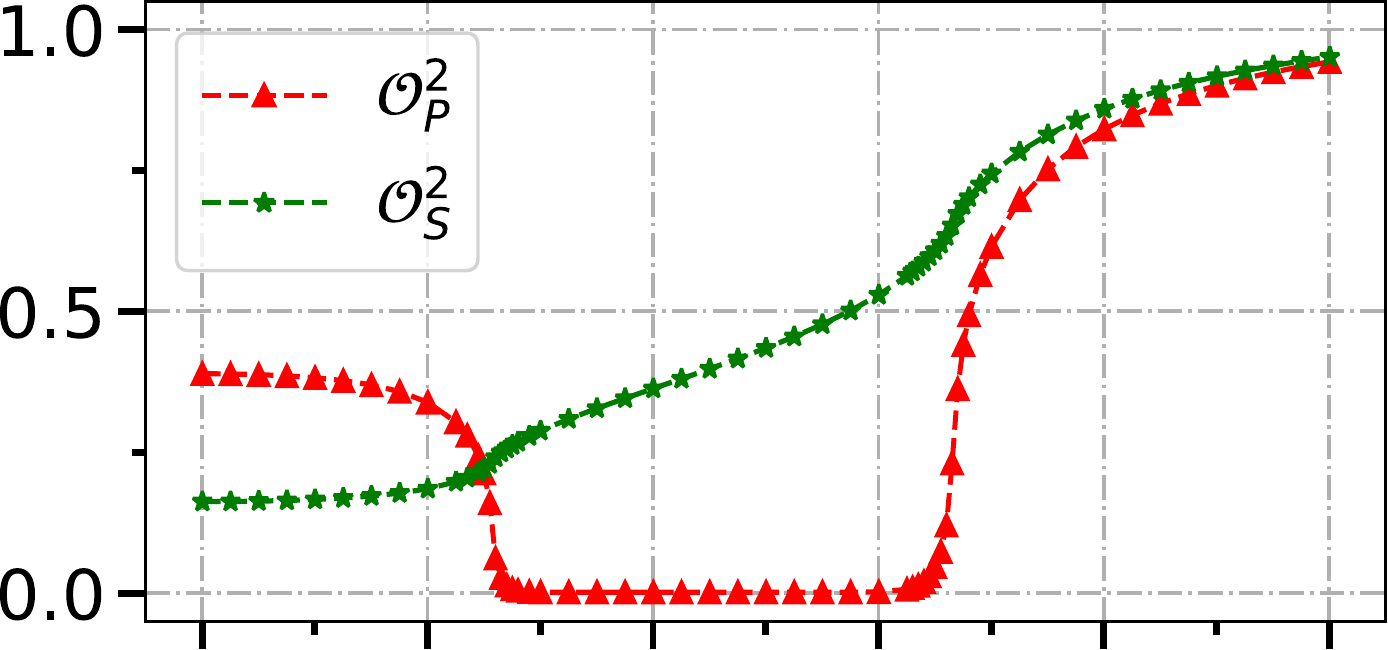}
	\vspace{1pt}
\end{minipage}

\begin{minipage}[t]{.05\linewidth}
\raisebox{3.1cm}[0cm][0cm]{(b)}
\end{minipage}
\begin{minipage}[t]{.7\linewidth}
	\includegraphics[width=.99\linewidth]{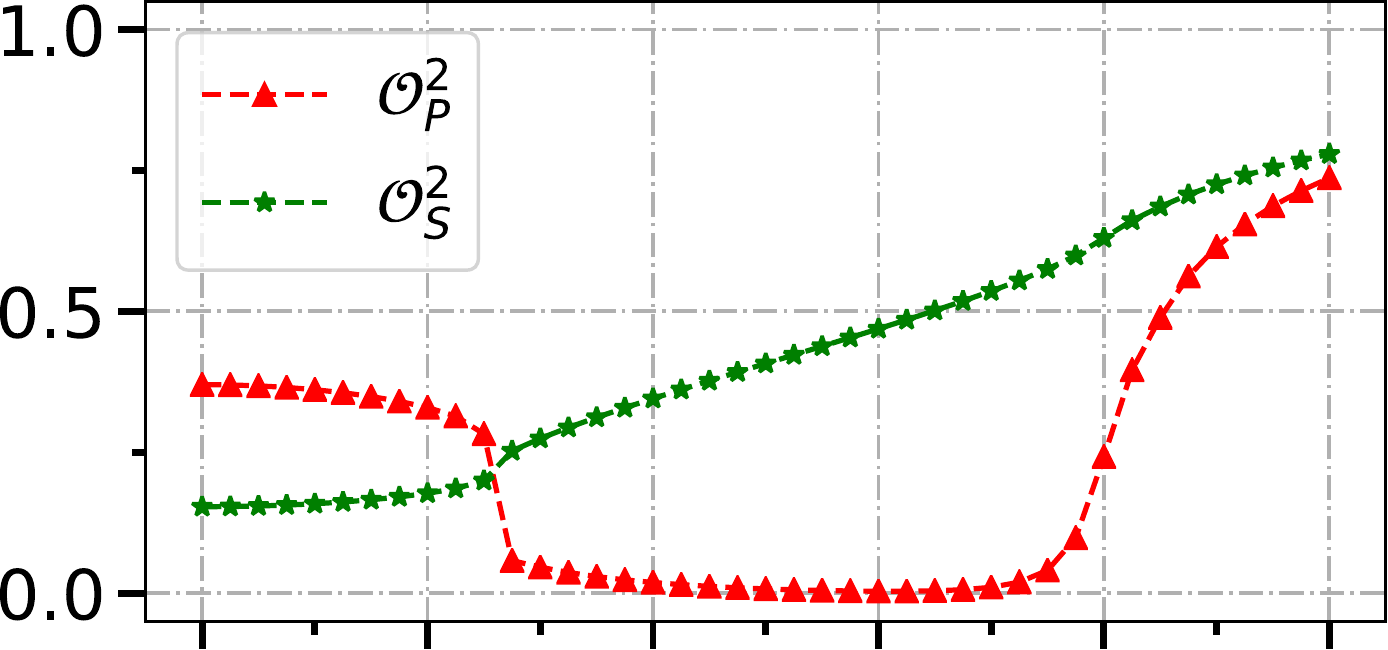}
	\vspace{1pt}
\end{minipage}

\begin{minipage}[t]{.05\linewidth}
\raisebox{3.1cm}[0cm][0cm]{(c)}
\end{minipage}
\begin{minipage}[t]{.7\linewidth}
	\includegraphics[width=.99\linewidth]{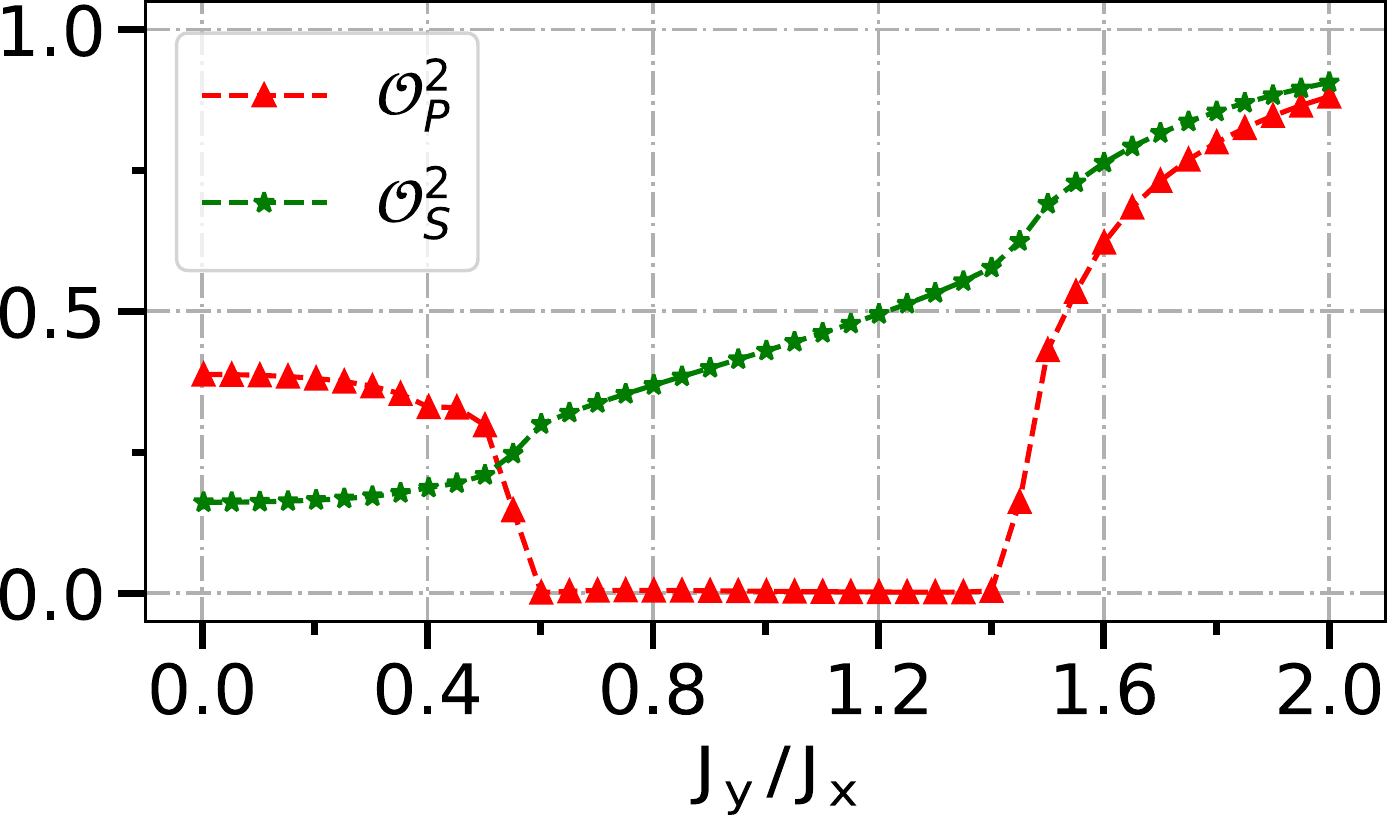}
\end{minipage}
\caption{Boundary-parity- and string-order of the QLM Eq.~(1) of the main text at $\mu=0$ for (a) two-leg ladder, (b) three-leg ladder, (c) four-leg ladder, with $L=100$ and $\chi=100$ (DMRG data).}
\label{fig:S_stringorder}
\end{figure}

For $\mu=0$ a symmetry protected topological phase~(SPT) emerges. This phase can be well characterized from properties of its entanglement spectrum, as shown in Fig.~\ref{fig:S_entanglement}~(a). The entanglement spectrum $\lambda_i$, which is the ordered sequence of Schmidt eigenvalues obtained for dividing the system into two parts along the central rung of the ladder, exhibits a twofold degeneracy in the SPT phase~\cite{Pollmann2010, Pollmann2012, Pollmann2012A}, as well as vanishing boundary parity and finite string order as shown in Fig.~\ref{fig:S_stringorder}~(a).

%%

%%%%%%%%%%%%%%%%%%%%%%%%%%%%%%%%%%%%%%%%%%%%%%%%%%%%%%%%%%%%%%%%%%%%%%%%%%%%%%%%%%%%%%%%%%
\section{Two-leg cylinder}

\begin{figure}[tb]
\centering
\includegraphics[width=0.7\linewidth]{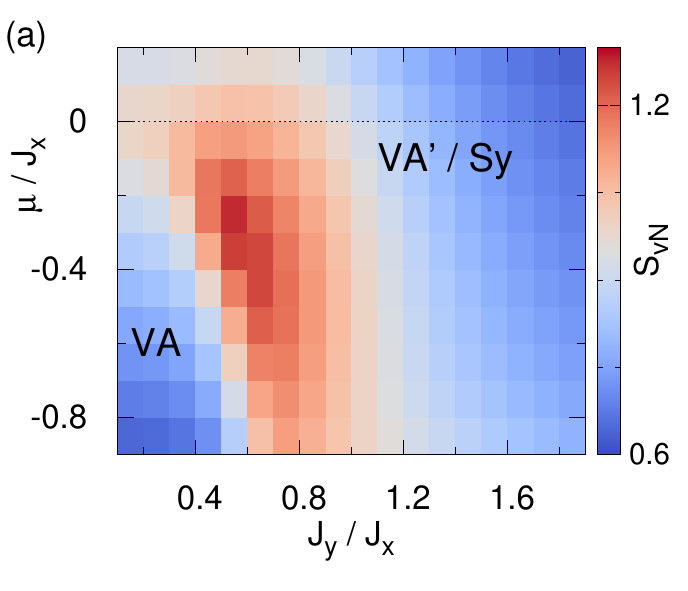}
\caption{
Schematic phase diagram of the two-leg cylinder model. Color codes depict the von-Neumann bipartite entanglement entropy $S_{vN}$ of the central rung (infinite DMRG simulation, $\chi=80$). 
\label{fig:S_pd_2torus} }
\end{figure}

For the two-leg cylinder we observe a drastically different ground-state phase diagram, as sketched in Fig.~\ref{fig:S_pd_2torus}. Here, the second order ring-change around the cylinder competes with local charge fluctuations in y-direction and, hence, the mechanism described in the main text to stabilize the distinct phases in the large $|\mu|\gg J_x, J_y$ limit is strongly affected.
In Fig.~\ref{fig:S_pd_2torus} we show numerical results for the phase diagram of the two-leg cylinder as function of $\mu$ and $J_y$. We only observe two clearly distinct phases. In the $J_y\ll J_x$ and $\mu<0$ regime, we observe a VA phase, which is characterized by an ordering of the link-spins in y-direction.
The VA phase which exhibits a quantum phase transition to a VA' like phase, which is adiabatically connected to the Sy phase. A distinct Sx phase is absent. 

%%%%%%%%%%%%%%%%%%%%%%%%%%%%%%%%%%%%%%%%%%%%%%%%%%%%%%%%%%%%%%%%%%%%%%%%%%%%%%%%%%%%%%%%%%
\section{Three- and four-leg ladder}

In Figs.~\ref{fig:S_entanglement}~(b) and (c) and ~\ref{fig:S_stringorder}~(b) and (c) we present additional data obtained from our DMRG calculations of the three and four-leg ladder systems at $\mu=0$. The exact twofold degeneracy in the entanglement spectrum is lost for more than two legs in the intermediate D phase. However, we observe the emergence of a distinct gap in the entanglement spectrum between a manifold of low lying and higher entanglement states. At the edge the string and parity order exhibit Haldane-like properties in the D phase~(Fig.~\ref{fig:S_stringorder}).

%%%%%%%%%%%%%%%%%%%%%%%%%%%%%%%%%%%%%%%%%%%%%%%%%%%%%%%%%%%%%%%%%%%%%%%%%%%%%%%%%%%%%%%%%%
\section{RK states}

\begin{figure}[tb]
\centering
\includegraphics[width=0.99\linewidth]{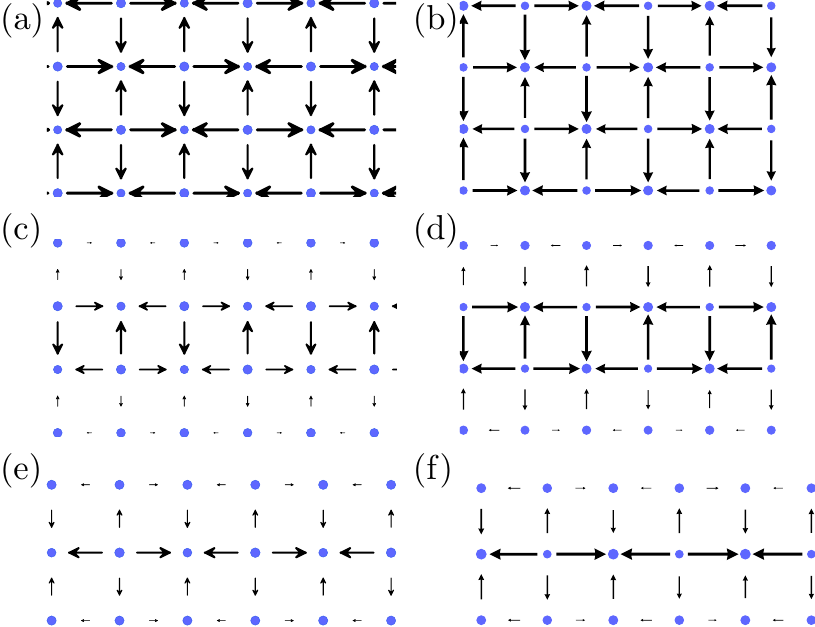}
\caption{Comparison of charge and spin configurations obtained from DMRG results for the intermediate phase ($J_x=J_y$, $\mu=0$)  (plots (a),(c), and (e)) to the RK state obtained by an equal amplitude overlap of all connected configurations (plots (b), (d), and (f)) for the four-leg  cylinder (a) and (b), the four-leg ladder (c) and (d), and the three-leg ladder (e) and (f)}
\label{fig:S_class_plt}
\end{figure}

In Fig.~\ref{fig:S_class_plt} we provide further comparisons to the classical RK states, obtained by a simple Metropolis sampling for the classical QLM analogue at infinite temperature, with the DMRG results of the intermediate phase in the QLM. For three- and four-leg ladders and four-leg cylinders the local spin and charge configurations compare very accurately.

%%%%%%%%%%%%%%%%%%%%%%%%%%%%%%%%%%%%%%%%%%%%%%%%%%%%%%%%%%%%%%%%%%%%%%%%%%%%%%%%%%%%%%%%%%
\section{Phase diagram}

In Tab.~\ref{tab:S_transitions} we summarize our results on the phase transition points for $\mu=0$ for the different systems analyzed in this work. Increasing the number of legs, the phase transition points behave non-monotonously. However, as a general trend the region of the intermediate phase expands and the transition points seem to approach a ratio $J_y^{c,1} J_y^{c,2} \sim 1$ (in units of $J_x$) as expected for the 2D limit with proper lattice rotation symmetry. 

\begin{table}[t]
\vspace{0.5cm}
\centering
\begin{tabular}{ c|c|c } 
  & $J_y^{c,1} \approx $ & $J_y^{c,2} \approx $ \\ 
 \hline
 Simplified 1D model & 0.2 & 1.1 \\ 
 two-leg ladder & 0.5 & 1.3 \\ 
 two-leg cylinder & - & - \\ 
 three-leg ladder & 0.5 & 1.6 \\ 
 four-leg ladder & 0.5 & 1.5 \\ 
 four-leg cylinder & 0.5 & 1.9 \\ 
 \hline
\end{tabular}
\vspace{0.3cm}
\caption{Estimated critical values of the exchange $J_y$ (in units of $J_x=1$) for the phase transition to the intermediate D phase for $\mu=0$.}
\label{tab:S_transitions}
\end{table}
%%

%%%%%%%%%%%%%%%%%%%%%%%%%%%%%%%%%%%%%%%%%%%%%%%%%%%%%%%%%%%%%%%%%%%%%%%%%%%%%%%%%%%%%%%%%%
\section{Extended data on the string tension}

\begin{figure*}[bt]
\centering
\includegraphics[width=0.99\linewidth]{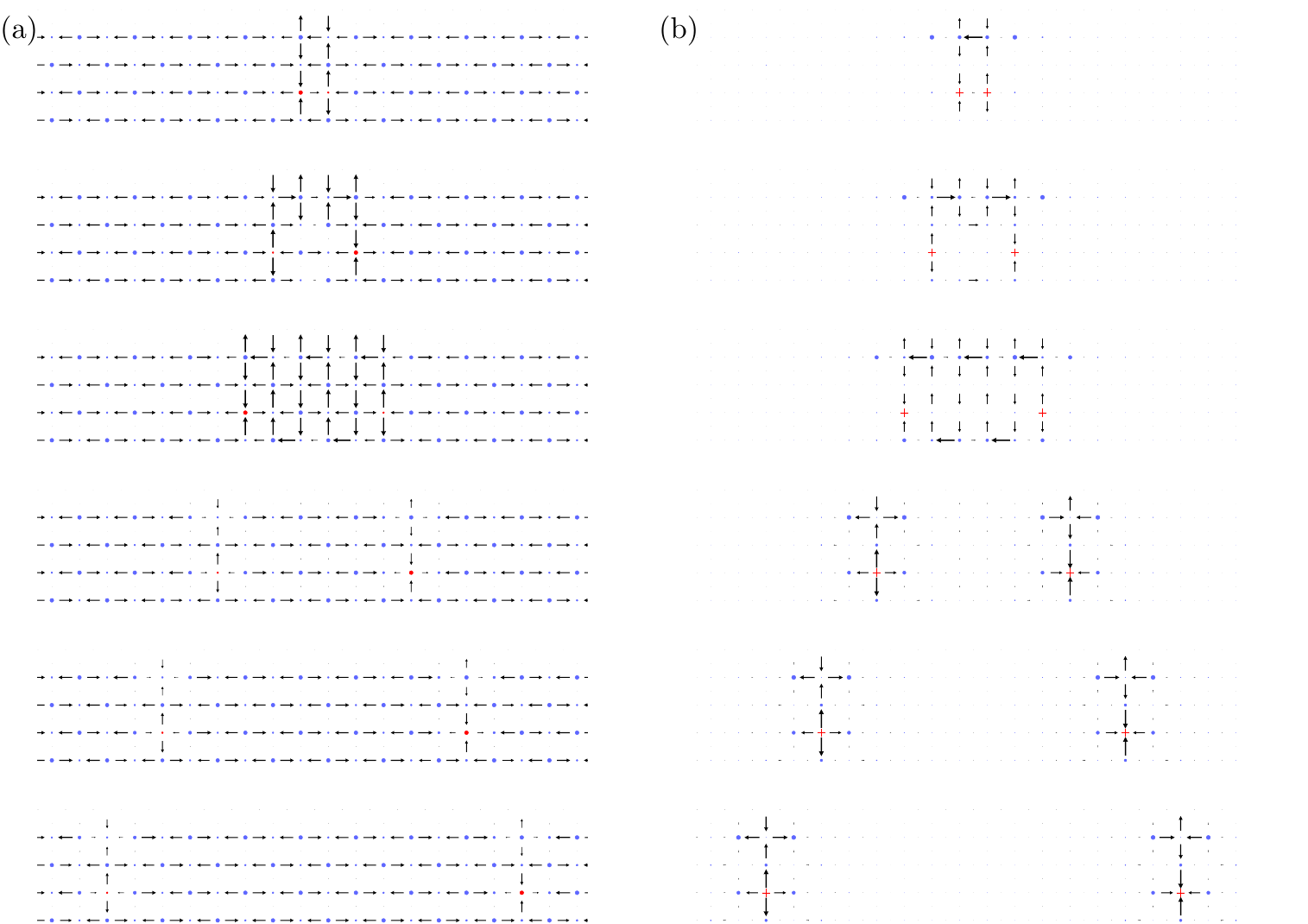}
\caption{(a) Charge and bond average configurations of two defects with distance $L_D=2$,$4$,$6$,$8$,$12$,$16$ sites. DMRG-data, $L=36$ rungs, $\mu=0.4 J_x$, $J_y / J_x = 0.4$ (b) Same as (a) but after substracting background without charges.}
\label{fig:S_pltST_jy0.4}
\end{figure*}

\begin{figure*}[bt]
\centering
\includegraphics[width=0.99\linewidth]{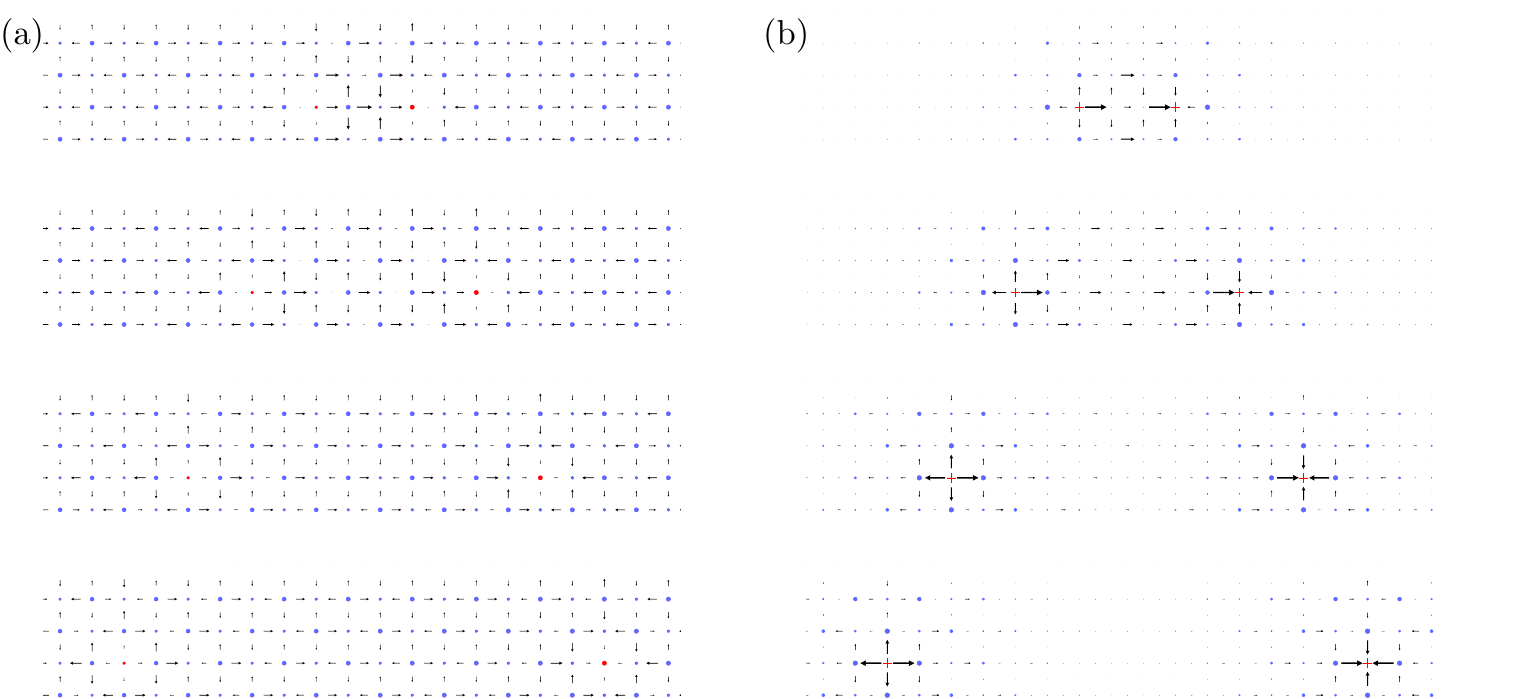}
\caption{(a) Charge and bond average configurations of two defects with distance $L_D=2$,$4$,$6$,$8$,$12$,$16$ sites. DMRG-data, $L=36$ rungs, $\mu=0.4 J_x$, $J_y / J_x = 1.0$ (b) Same as (a) but after substracting background without charges.}
\label{fig:S_pltST_jy1.0}
\end{figure*}

\begin{figure*}[bt]
\centering
\includegraphics[width=0.99\linewidth]{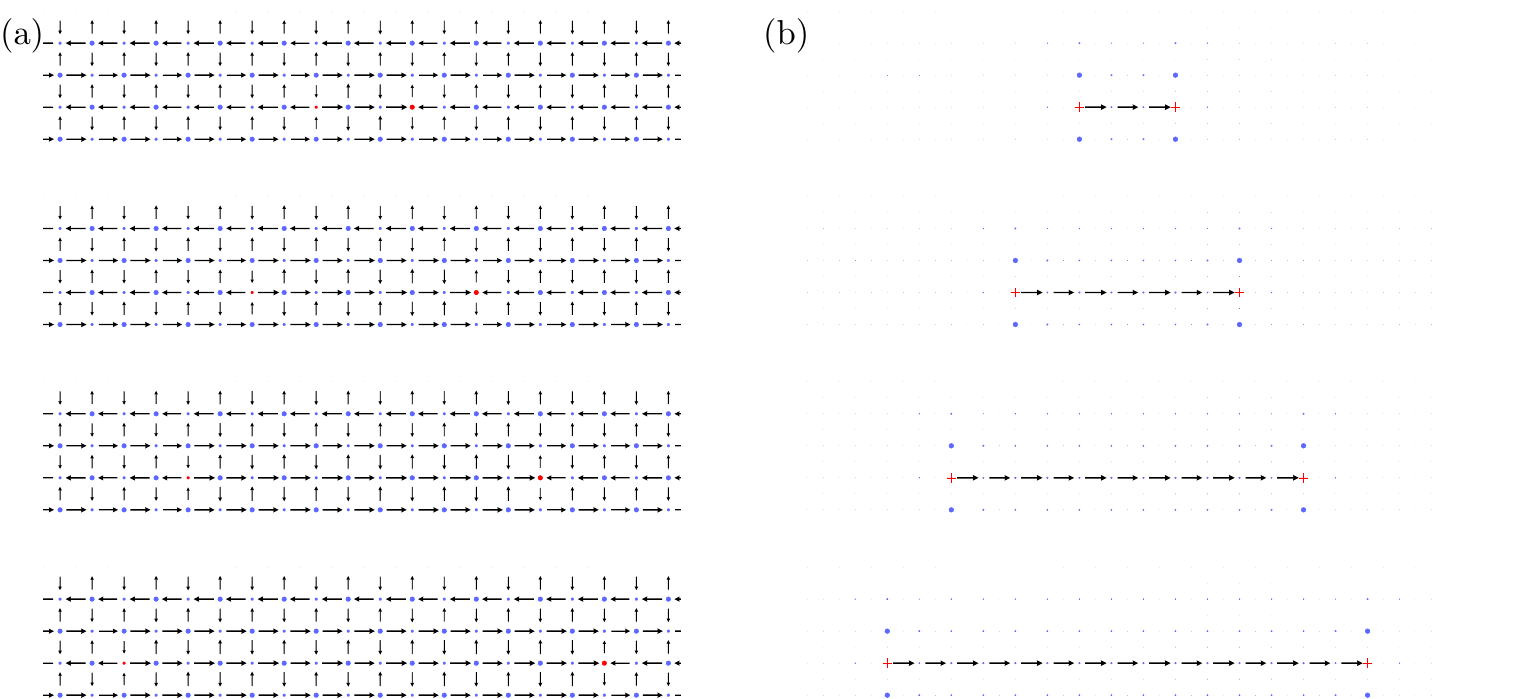}
\caption{(a) Charge and bond average configurations of two defects with distance $L_D=2$,$4$,$6$,$8$,$12$,$16$ sites. DMRG-data, $L=36$ rungs, $\mu=0.4 J_x$, $J_y / J_x = 1.8$ (b) Same as (a) but after substracting background without charges.}
\label{fig:S_pltST_jy1.8}
\end{figure*}

In Figs.~\ref{fig:S_pltST_jy0.4}, \ref{fig:S_pltST_jy1.0}, \ref{fig:S_pltST_jy1.8} we finally detail the DMRG results for the calculation of the string-formation and tension between defects as shown in Figs.~(4) and (5) of the main text. 
We depict three examples from the Sy (Fig.~\ref{fig:S_pltST_jy0.4}), D phase (Fig.~\ref{fig:S_pltST_jy1.0}) and Sx (Fig.~\ref{fig:S_pltST_jy1.8}) phases without and with subtraction of the background charge and spin configurations. 

%%%%%%%%%%%%%%%%%%%%%%%%%%%%%%%%%%%%%%%%%%%%%%%%%%%%%%%%%%%%%%%%%%%%%%%%%%%%%%%%%%%%%%%%%%
\bibliography{references}